# Classification of La$^{3+}$ and Gd$^{3+}$ rare earth ions using surface-enhanced Raman scattering

Hao Jin[1], Tamitake Itoh[2] and Yuko S. Yamamoto[1]*

[1] School of Materials Science, Japan Advanced Institute of Science and Technology (JAIST), Nomi, Ishikawa 923-1292, Japan

[2] Nano-Bioanalysis Research Group, Health Research Institute, National Institute of Advanced Industrial Science and Technology (AIST), Takamatsu, Kagawa 761-0395, Japan

*Corresponding author: yamayulab@gmail.com

**Abstract**

In this study, surface-enhanced Raman scattering (SERS) spectra of different rare earth (RE) ion-citrate complexes were investigated for the first time for the qualitative classification of $RE^{3+}$ ions. With the addition of $RE^{3+}$ ions to citrate-capped silver nanoparticles in aqueous solutions, the Raman signals of RE-citrate complexes were enhanced, and characteristic peaks appeared near 1065 cm$^{-1}$ and 1315 cm$^{-1}$. The $I_{1065}/I_{1315}$ ratios of La-citrate and Gd-citrate were approximately 1 and 0.5, respectively. Thus, different $RE^{3+}$ ions were classified based on the ratio of characteristic SERS peaks near 1065 cm$^{-1}$ and 1315 cm$^{-1}$. In addition, the effects of $RE^{3+}$ ions in the RE-citrate complexes were analyzed based on density functional theory (DFT) calculations. Calculation results show that these characteristic peaks are attributed to the coordination of carboxyl and hydroxyl groups of citrates with the $RE^{3+}$ ions, suggesting that these are spin-related bands of the RE-citrate complexes.

**Introduction**

Owing to their unique 4f electronic configuration, rare earth (RE) elements show similar chemical properties, although they have unique physical properties, e.g. spin, electric, and optical characteristics [1]. When $RE^{3+}$ ions are combined with small molecules to form complexes, these complexes also exhibit such properties, showing differences in luminescence [2-4] and magnetism [3-4], therefore, RE ion-molecule complexes have a wide range of applications such as phosphors[5], contrast agents[6], single-molecule magnets[7], and bio-probes for theranostics [8]. They can also be used to produce light-matter interfaces at the quantum level to archive quantum calculations[9]. Standard analytical techniques like NMR spectroscopy, ESR spectroscopy, fluorescence spectroscopy, and multi-modality theranostics are powerful methods for studying the function and action of RE ion-molecule complexes[2-10]; however, these methods are inextricably linked to complex sample preparation processes and it is difficult to measure specific characteristic signals when the sample is mixed with different $RE^{3+}$ ions[10]. Raman spectroscopy is a nondestructive analytical detection technique by which one can obtain stable molecular fingerprint features, such as molecular structure information[11] and spin information[12][13]. The use of noble metal nanostructures as Raman substrates significantly enhances the Raman signal, enabling the detection of analytes at low concentrations ($<10^{-5}$M) even in single-molecule detection in aqueous solutions[14]. This phenomenon is called surface-enhanced Raman scattering (SERS)[15-17]. In the SERS process, the enhancement of the Raman signal is mainly due to localized surface plasmon resonance (LSPR), which is generated by collective oscillations of conduction electrons in noble metal nanoparticles[18][19]. Furthermore, charge transfer between the noble metal and molecules attached to the metal is also an important mechanism of SERS enhancement, known as the chemical mechanism[20]. Thus, SERS has the potential to be used to study RE-molecule complex systems in terms of $RE^{3+}$ ion detection, functional characterization of complexes, and optical signal modulation.

However, SERS studies of $RE^{3+}$ ions remain challenging because of the similarities in the chemical properties of RE elements. Researches on $RE^{3+}$ ions in the field of SERS have focused on studying the effects of individual $RE^{3+}$ ions rather than identifying RE elements, such as adding $Nd^{3+}$ ions to study the enhancement mechanism of semiconductor SERS substrates[21], fabricating Gd-fluorescent complexes for MRI-SERS multimodal detection[22], and exploring the fluorescence enhancement mechanisms of individual RE elements[23][24]. SERS studies of RE's effect remain limited owing to the difficulty of $RE^{3+}$ ion measurements; for example, the addition of $Gd^{3+}$ ions to the SERS substrate produces a stronger SERS signal[25], but it is not certain whether the addition of other $RE^{3+}$ ions also results in a stronger SERS signal. Moreover, the fluorescence background of some $RE^{3+}$ ions e.g. $Pr^{3+}$, $Sm^{3+}$, and $Eu^{3+}$ [26], may interfere with the SERS signals; therefore, extracting the characteristic signals which are related to the special physical properties of $RE^{3+}$ ions from the SERS spectra of RE-molecule complexes is also challenging. Moreover, the mechanism by which spin differences affect the SERS spectra is not clear. Although some previous studies [12][13] have achieved building the spin effects model for SERS/Raman of metmyoglobin complex including $Fe^{2+}$ or $Fe^{3+}$ ions, this process appears to be influenced by the coordination conditions of the ligand and metmyoglobin, therefore, the core mechanisms underlying the spin properties should be further investigated.

In this study, we selected two non-fluorescent $RE^{3+}$ ions, $La^{3+}$ and $Gd^{3+}$, to investigate the spin effect of $RE^{3+}$ ions on the enhancement and characteristic differences of SERS spectra. RE-citrate complexes were evaluated by SERS using citrate-capped silver nanoparticles (citrate@AgNPs), which are well-known SERS substrates for application in the visible-light region[27] then we found the SERS spectral differences that enabled the identification of $La^{3+}$ and $Gd^{3+}$. These differences in SERS spectra may be attributed to the spin state differences of $La^{3+}$ and $Gd^{3+}$ because the chemical properties of $La^{3+}$ and $Gd^{3+}$ are similar but $La^{3+}$ has no spin, and $Gd^{3+}$ has the highest spin state due to their special 4f electron configurations, i.e., 0 and 7 electrons in 4f orbital, respectively. These results will be beneficial for future studies on the identification of RE elements by SERS, and potentially provide new insights into the effect of spin state differences in RE elements on SERS spectra.

**Experiments**

**Solvent and chemicals.** Ultrapure water produced on a Direct-Q® UV 3 system (Millipore, USA) was used as the solvent. A $HNO_3$ (1.38g/mL, Kanto Chemical Co., Inc, Japan) solution was used to prepare 1 M $HNO_3$ solution. 0.1 M $RE(NO_3)_3$ (RE: La, Gd) solutions were obtained by dissolving an appropriate amount of RE oxides, i.e., $La_2O_3$ and $Gd_2O_3$ (≥99.9%, FUJIFILM Wako Pure Chemical Corporation, Japan) using 1 M $HNO_3$ solution by heating at 100 °C, and $10^{-3}$M $RE(NO_3)_3$ solutions were obtained by diluting 0.1M $RE(NO_3)_3$ solutions with ultrapure water. Citrate@AgNPs were prepared using the Lee & Meisel method[28]. Briefly, 0.03 g of silver nitrate (FUJIFILM Wako Pure Chemical Corporation, Japan) and 0.03 g of trisodium citrate (FUJIFILM Wako Pure Chemical Corporation, Japan) were added to 150 mL of boiling ultrapure water, heated, and stirred at 150 °C in the oil bath for an hour. When the solution turned gray-green, heating was stopped, and the solution was allowed to cool naturally to room temperature to obtain a solution of citrate@AgNPs. The citrate@AgNPs solution was stored away from light in a refrigerator at 4°C.

**Characterization.** Raman samples were prepared by heating and dissolving 0.01 g of trisodium citrate in 50 mL of water, and then mixing trisodium citrate solution (7.75 ×$10^{-4}$M, 1 mL), and RE(NO$_3$)$_3$ solution ($10^{-3}$M, 50 μL) to prepare a 5×$10^{-5}$ M RE-citrate solution. Samples for SERS measurements were obtained by adding 50 μL of $10^{-3}$M RE(NO$_3$)$_3$ solutions to separate 1 mL solutions of citrate@AgNPs. The concentrations of RE-citrate complexes were 5×$10^{-5}$M for each RE-citrate@AgNPs sample. The original citrate@AgNP colloidal solution without any ions or salt addition was also used for the SERS measurement as a blank experiment to analyze the potential effect of RE$^{3+}$ ions on the SERS substrate. Note that the original citrate@AgNP solution is considered to contain independent AgNPs without aggregations, therefore SERS signal obtained was weaker than other samples due to the lacks of SERS hotspots. All samples' pH were measured using pH test strips (REF 92150 and REF 92140; MACHERY-NAGEL, USA), and their pH was approximately 7.0. All sample solutions were stored at room temperature for 12 hours prior to spectroscopic measurements.

For UV-Vis measurements, ultrapure water was added to each SERS sample to obtain a 10-fold dilution of the sample solutions then the UV-Vis measurements were performed using a V-770 spectrometer (JASCO, Japan) with a polystyrene cuvette having a light path of 1 cm. Ultrapure water was used as the blank to measure the UV-Vis spectra.

For Raman and SERS measurements, a clean soda glass capillary was used to hold the sample solution for Raman and SERS spectroscopic measurements which were performed using a T64000 Raman spectrometer (Horiba Scientific, Japan). All Raman spectra were collected using lasers at wavelengths of 488 nm and 532 nm with a power level of 50 mW and a 100× objective (0.75. N.A.) that when characterizing the samples with 488 nm and 532 nm lasers, the energy power at the sampling point will be 101 and 85 mW $\mu m^{-2}$, respectively. The exposure time of Raman and SERS measurements was 30 seconds and the accumulation was 2. Raman shift were corrected by indene.

To perform spectra fitting and baseline subtraction for all experimental spectra, the OriginPro 2022 (OriginLab Corporation, U.S.A.) software was used. The Savitzky-Golay method was used for noise reduction twice with setting as 5 points of windows and polynomial order was 1. The cubic function was chosen for baseline subtracting.

**Computational Methods and Details**

The SERS simulations of RE-citrate complexes on the silver surface were performed using the following simplified model: according to Ref 29, the structures of RE-citrate complexes were modeled as RECit$^-$ based on the structure of trisodium citrate which was downloaded from the Cambridge Crystallographic Data Centre (CCDC,

http://www.ccdc.cam.ac.uk). An appropriate size Ag surface formed by 11 Ag atoms was cut from a bulk Ag crystal along the (111) plane[30] to mimic the facets of the silver surface, whose size is close to RECit$^-$.

Further, geometry optimization and vibrational spectra calculations were performed based on density functional theory (DFT) using Gaussian 16 software. The commonly used B3LYP functional is not suitable for describing transition metals[31]. Therefore, the PBE0 functional[32] was used with the addition of DFT-D3 dispersion correction[33] for all calculations to better describe the 4f electron configuration of RE elements[34][35]. The solvent was considered as water using SMD model to match the experimental parameters. The aug-cc-pvdz [36] basis set was adopted C, H, O, and Na. In contrast, Ag and RE elements were described by the def2svpd[37] and def2tzvpd[38][39] basis sets, respectively, and all associated pseudopotentials[40]. The scaling factors for PBE0-D3/aug-cc-pvdz and PBE0-D3/def2tzvpd; were 0.956 and 0.971, respectively, as obtained from the literatures[41][42]. The calculation results of SERS spectra in the region of 1000 − 1700 cm$^{-1}$ were scaled by scaling factor 0.956 and in the region of 1000 − 1250 cm$^{-1}$, spectra were scaled by scaling factor 0.971 again to scale RE$^{3+}$ ions' effect. The widths of all simulated SERS spectra were set to 50 cm$^{-1}$ to approximate the experimental spectra. All energy calculations of the RE-citrate complexes were performed at the PBE0-D3/def2tzvpp[43] level. All calculation results were analyzed using the Multimfn software[44], including Raman analysis and spin analysis. Gaussian 16 directly calculates the Raman activity of each vibrational mode. The distribution of spin density can be described by spin population analysis by using Multimfn. The Visual Molecular Dynamics (VMD) software [45] was used to observe the structure and spin distribution.

**Results and discussion**

**Raman and SERS measurements of RE-citrate complexes (RE: La, Gd).**

Figure 1(a) shows a schematic representation of the interaction of $RE^{3+}$ (RE: La, Gd) ions with the surfaces of the citrate@AgNPs. A citrate@AgNPs solution was synthesized by Lee & Meisel method using citrate molecules as a reducing agent[28]; thus, all the AgNP surfaces were covered by citrate molecules[27][46][47]. Therefore, in this manuscript, we refer to AgNPs as citrate-capped AgNPs, namely citrate@AgNPs. Figure 1(b) shows the differences between the UV-visible spectra of citrate@AgNP colloidal solutions collected before and after the addition of the different $RE(NO_3)_3$ aqueous solutions. In this process, $RE^{3+}$ ions are adsorbed by the citrate on the silver surface and coordinate with the hydroxyl and carboxyl groups of citrate[29][48] to form a 1:1 RE-citrate complex[49] at pH=7.0 as main structure as $RECit^-$[29], possibly resulting in the fabrication of AgNPs aggregates. Usually, the junction of AgNPs aggregates generates a strong electromagnetic field under laser excitation, namely hotspot[50][51]; therefore, we can expect a much stronger Raman signal of RE-citrate complexes at hotspots between the AgNPs.

As shown in Figure 1(b), on the addition of different $RE^{3+}$ ions to separate citrate@AgNP colloidal solutions, the UV-visible absorption spectra of the mixtures do not show the change in their peak positions; however, the peak heights are decreased, showing that AgNPs aggregates are fabricated upon addition of $RE^{3+}$ ions. The absorption peaks of the three samples were all around 450 nm, indicating that the addition of $RE^{3+}$ ions did not affect the location of the main absorption peaks, and therefore in this study, we chose 488 nm and 532 nm laser light, respectively, as Raman and SERS laser light sources for higher SERS enhancement. A measurement system for Raman and SERS spectra is shown in Figure 1(c), where the data obtained with and without citrate@AgNPs samples at the same measurement parameters are comparable for evaluating the enhancement effect, laser wavelength effects and ions' effects.

Figure 2 shows representative Raman and SERS spectra of the La-citrate and Gd-citrate sample solutions excited at 488 nm or 532 nm, respectively. The Raman spectra of the samples were collected without citrate@AgNPs in the sample inserted into the glass capillary, while the SERS spectra of the samples were collected with citrate@AgNPs. Without citrate@AgNP, the Raman signals of the RE-citrate solutions at the same concentration were very weak with no significant signal peaks. Upon addition of citrate@AgNPs in the RE-citrate solution, the Raman signal collected was stronger than that without citrate@AgNPs, indicating that the existence of citrate@AgNPs enhanced the Raman signal of the RE-citrate solution, i.e. SERS occurred. For the peak assignments, DFT calculations were introduced along with the results of previous analyses of the literatures [27][52] and were summarized in Table 1 and 2.

In the SERS spectrum of Figure 2(a), there were several inconspicuous peaks between 690 and 900 cm$^{-1}$, such as peaks at 697, 737, and 775 cm$^{-1}$, which corresponded to different $\delta(COO^-)$, $\delta(COO^-) + \gamma(COO^-)$ and $\delta(COO^-) + \gamma(COO^-)$. Some inconspicuous peaks near 825 cm$^{-1}$ that corresponded to $\nu(CCCC-O)$ with lower relative intensities than the peak at 775 cm$^{-1}$. The peaks at 737 and 775 cm$^{-1}$ exhibited similar relative intensities and were both lower than the peak at 697 cm$^{-1}$. While other significant Raman peaks appeared at 945, 1066, 1314, and 1496 cm$^{-1}$, which can in turn be assigned as $\nu(C-COO^-)$, $\nu(C-O-La) + \gamma(CH_2)$, $\nu_{sym}(COO^-) + \delta(CH_2)$, and $\nu_{asym}(COO^-) + \gamma(CH_2)$.

Figure 2(b) shows the Raman and SERS spectra of the La-citrate solution excited at 532 nm. The Raman signal of the La-citrate solution collected using the 532 nm laser has one significant Raman peak at 1644 cm$^{-1}$, potentially exhibiting acetoacetic acid Raman peak due to the thermal decomposition of citrate[27][53]. In the SERS spectrum, the slight difference of some vibration bands compared to the La-citrate's SERS spectrum excited by the 488 nm laser may be related to the signal noise and the influence of laser wavelength. The peaks at 697, 736, and 769 cm$^{-1}$ were corresponded to $\delta(COO^-)$ and different $\delta(COO^-) + \gamma(COO^-)$. The peak at 822 cm$^{-1}$ was corresponded to $\nu(CCCC-O)$ and its relative intensity was close to that of the peak at 937 cm$^{-1}$ but lower than that of peaks at 697, 736, and 769 cm$^{-1}$. The significant Raman peaks appeared at 937, 1065, and 1309 cm$^{-1}$, which can in turn be assigned as $\nu(C-COO^-) + \delta(CH_2)$, $\nu(C-O-La) + \gamma(CH_2)$ and $\nu_{sym}(COO^-) + \delta(CH_2)$. respectively. The band contains $\nu_{asym}(COO^-) + \gamma(CH_2)$ was widely and detected around 1500 cm$^{-1}$.

Figure 2(c) shows representative Raman and SERS spectra of the Gd-citrate solution excited at 488 nm. Similar to the case of La-citrate, when Gd-citrate was used without citrate@AgNPs, the Raman signal was very weak. However, when Gd$^{3+}$ ions were added to citrate@ AgNPs to form Gd-citrate on the surface of the AgNPs, the SERS signal was occurred. In the SERS spectrum, several inconspicuous peaks appeared at 697, 742, 776, and 822 cm$^{-1}$, corresponding $\delta(COO^-)$, $\delta(COO^-) + \gamma(COO^-)$, $\delta(COO^-) + \gamma(COO^-)$ and $\nu(CCCC-O)$. The relative intensities of these peaks were lower than the intensities of the peaks near 937 cm$^{-1}$. And the relative intensities of peaks that appeared at 742 and 776 cm$^{-1}$ were similar. Four significant characteristic peaks were observed at 937, 1066, 1315, and 1495 cm$^{-1}$, which can in turn be assigned as $\nu$(C-COO$^-$) + $\delta$(CH$_2$), $\nu$(C-O-Gd) + $\gamma$(CH$_2$), $\nu_{sym}$(COO$^-$) + $\delta$(CH$_2$), and $\nu_{asym}$(COO$^-$) + $\gamma$(CH$_2$), respectively. The small differences in positions of each vibration band can be explained by RE$^{3+}$ ion's differences.

Figure 2(d) shows the Raman and SERS spectra of Gd-citrate excited at 532 nm. One significant Raman peak was observed at 1647 cm$^{-1}$ and this peak was not observed in the SERS spectrum significantly. In the SERS spectrum, the peaks between 690 and 900 cm$^{-1}$ were observed at 696, 737, 772, and 818 cm$^{-1}$, respectively. In addition, the intensity of the peak at 737 cm$^{-1}$ was higher than that of the peak at 772 cm$^{-1}$ since these two peaks corresponded to different $\delta$(COO$^-$) + $\gamma$(COO$^-$). The peak at 818 cm$^{-1}$ was assigned as $\nu$(CCCC-O). Compared with the peak at 937 cm$^{-1}$, the relative intensities of these three peaks were lower. Other significant SERS characteristic peaks over this region were observed at 937, 1065, 1313, and 1495 cm$^{-1}$, assigned as $\nu$(C-COO$^-$) + $\delta$(CH$_2$), $\nu$(C-O-Gd) + $\gamma$(CH$_2$), $\nu_{sym}$(COO$^-$) + $\delta$(CH$_2$), and $\nu_{asym}$(COO$^-$) + $\gamma$(CH$_2$), respectively.

Notably, compared with the evaluation of the enhancement in the citrate solution without RE$^{3+}$ ions (Figure S4), the Raman signals of citrate solutions with RE$^{3+}$ ions were significantly enhanced by citrate@AgNPs when the target samples contained RE$^{3+}$ ions.

Moreover, when comparing the Raman spectrum of citrate with the same concentration and the Raman spectrum of citrate@AgNPs without adding any ions or salts (SERS spectrum of citrate), the Raman signal was not significantly enhanced. This is due to a lack of aggregation of sliver nanoparticles, and not generating enough hotspots.

**Classification of $RE^{3+}$ ions by SERS.**

Further analysis of the experimental SERS spectra in the region of 600 − 1800 $cm^{-1}$ was shown in Figure 3 (a) and (b). The SERS intensity of RE-citrates was several times higher than that of citrate due to the hotspot regions formed after adding $RE^{3+}$ ions. The characteristic peak positions in the SERS spectrum obtained by RE-citrate were close to each other corresponding to the same vibrational model that analyzed by DFT simulation. Thus, approximate position classifications were made based on vibration bands of SERS characteristic peaks, which were marked with a black dotted line. In comparison with the SERS spectrum without $RE^{3+}$ ions (blue line), the characteristic peaks in the SERS spectra with $La^{3+}$ ions (black line) and $Gd^{3+}$ ions (red line), inconspicuous peaks were appeared in the region of 700 − 825 $cm^{-1}$, and significant peaks were at 940, 1065, 1315 and 1495 $cm^{-1}$, respectively. We mainly analyzed the relative intensities of peaks in the interval from 1000 $cm^{-1}$ to 1800 $cm^{-1}$ in Figure 3. We observed that the relationship between the relative intensities of the peaks near 1065 and 1315 $cm^{-1}$ was determined by the type of RE$^{3+}$ ions, independent of the laser wavelength. For La-citrate, the relative intensities of these two SERS peaks were almost identical. On the other hand, for Gd-citrate, the intensity of the SERS peak near 1065 cm$^{-1}$ was almost half of the SERS peak intensity at 1315 cm$^{-1}$. However, when the laser wavelength was changed, the peaks' relative intensities of peaks at 1315 and 1495 cm$^{-1}$ in SERS spectra of La-citrate and Gd-citrate were significantly changed. Only when the laser wavelength was 532 nm, the intensities of peaks near 1315 cm$^{-1}$ were higher than the intensities of peaks near 1495 cm$^{-1}$. Therefore, we focused on the analysis of these three SERS peaks that corresponding the hydroxyl and two carboxyl groups of citrates coordinated with RE$^{3+}$.

Figures 4(a) and (b) show the changes in the SERS spectra in the region of 1000 − 1400 cm$^{-1}$ upon the addition of La$^{3+}$ and Gd$^{3+}$ ions to citrate@AgNPs under laser excitation at different wavelengths. With the addition of La$^{3+}$ ions, the intensities of the characteristic peaks around 1065 cm$^{-1}$ remain basically the same as the intensities of the characteristic peaks around 1315 cm$^{-1}$. When Gd$^{3+}$ ions were added, the intensities of the characteristic peaks at approximately 1065 cm$^{-1}$ were significantly weaker than those at approximately 1315 cm$^{-1}$, and this relative SERS intensity relationship was not affected by the laser wavelength.

Figure 4(c) shows the relative intensity ratios of these two characteristic peaks around 1065 and 1315 cm$^{-1}$ under laser excitation at different wavelengths. In this figure, the I$_{1065}$/I$_{1315}$ ratios of La-citrate and Gd-citrate are approximately 1 and 0.5, respectively, enabling us to classify La$^{3+}$ and Gd$^{3+}$ ions by SERS measurements.

In comparison to Figures 4(a)-(c), Figures 4(d) and (e) show a negative description of the changes in the SERS spectra in the region of 1200 − 1800 cm$^{-1}$ for the addition of La$^{3+}$ and Gd$^{3+}$ ions to citrate@AgNPs under laser excitation at different wavelengths. There are two characteristic peaks in this region, at around 1315 cm$^{-1}$ and 1 cm$^{-1}$, when the RE-citrate@AgNPs were excited at 488 nm and 532 nm. Moreover, the relative intensity ratios I$_{1495}$/I$_{1315}$ of RE-citrate shown in Figure 4(f) did not exhibit a clear difference between the La$^{3+}$ and Gd$^{3+}$ ions.

**Analysis of spin effects of RE elements in SERS spectra**

Here, we discussed how different RE$^{3+}$ ions, La$^{3+}$ and Gd$^{3+}$, can be classified using SERS measurements, even though these ions have similar chemical properties and outer electron configurations. We referred to previous studies[12][13] and found that the differences between the intensity ratios of these SERS characteristic peaks are related to the spin state difference of the ions. Theoretically, owing to the difference in electron configuration of La$^{3+}$ (electron configuration: [Xe]4f$^0$, no spin) and Gd$^{3+}$ (electron configuration: [Xe]4f$^7$, highest spin) ions produce a large spin difference. When the RE$^{3+}$ ion and citrate coordinate to form a complex, the 4f orbital is the inner orbital, and such a difference can still be maintained. Citrate molecule has one hydroxyl (-OH) group and three carboxyl (COO$^-$) groups. For the RE-citrate complexes, the (C-O-RE) and (COO-RE) bands were generated by the coordination of RE$^{3+}$ ions with the hydroxyl and two carboxyl groups of citrates. These two bands present the difference, especially the spin difference between La$^{3+}$ and Gd$^{3+}$ ions and were confirmed that corresponding the SERS peaks around 1065 cm$^{-1}$, and peaks around 1315 and 1495 cm$^{-1}$. Thus, combining the results of Figure 4(c), mainly owing to the spin state difference of La$^{3+}$ and Gd$^{3+}$ ions, the relationship of the difference between these two bands around 1065 cm$^{-1}$ and 1315 cm$^{-1}$ in the SERS spectrum can be used for the classification of La$^{3+}$ and Gd$^{3+}$ ions.

We also used DFT calculation results to analyze the sources of the differences between these two vibration bands. To obtain simulated Raman spectra, the Raman activity of each vibrational mode should be converted to the Raman intensity of each vibrational mode by the following equation[54].

$$I_i = \frac{C(v_0 - v_i)^4 S_i}{v_i B_i}; B_i = 1 - \exp\left(-\frac{hvc_i}{k_B T}\right) \qquad (1)$$

where *i* denotes vibrational mode, C is the normalization factor and can be arbitrarily chosen, *v* and *v₀* correspond to vibrational frequency and frequency of incident light, respectively. *S* is Raman activity calculated by Gaussian 16. h, c, k$_B$, and T are Planck constant, light speed, Boltzmann constant, and temperature, respectively. In the present work, the value of $v_0$ and T were chosen to be 532 nm and 298.15K, respectively, for the better simulation. On another hand, Raman scattering is produced by the induced dipole moment, which is the change of the polarizability around the equilibrium nuclear coordinates[11]. In general, larger electronic density results in greater resistance to polarization. In our present RE-complex systems, the vibration modes of each RE-complex are similar, but the $RE^{3+}$ ions have a huge difference in the number of single electrons in the 4f orbital, which affects the induced dipole moment, and thus has different intensity differences for the same Raman peak. It suggests that the spin density can describe the distribution of single electrons which indirectly shows the condition of electron density difference[55].

Figure 5(a) shows the structure and spin population of the La-citrate complex. This complex does not contain single electrons, and thus its spin density is zero. Figure 5(b) shows a comparison of the scaled simulated SERS spectrum and the experimental spectrum under 532 nm excitation of La-citrate. The main characteristic peaks of the (C-O-La) and (COO$^-$) bands were well matched, the intensity of the $\nu$(C-O-La) was close to that of the $\nu_{sym}$(COO$^-$), while the intensity of the $\nu_{sym}$(COO$^-$) was stronger than that of $\nu_{asym}$(COO$^-$) both in simulated and experimental spectra; thus, the spectral shapes were also basically the same. There are some discrepancies between the simulated and experimental spectra, which may be due to the mixture of binding modes of actual citrate, resulting in the inability of a single simplified model to simulate the full details[47] and the effects of mixed basis sets. These discrepancies were acceptable after scaling using suitable scaling factors[39][40].

Figure 5(c) shows the structure and spin population of Gd-citrate. The spin density of the carboxyl and hydroxyl groups coordinated with the $Gd^{3+}$ ion was not 0, which was because the $Gd^{3+}$ ion has seven unpaired electrons in the 4f orbital. Thus, combining the results of La-citrate, the (C-O-RE) and the ($COO^-$) bands were affected by the $RE^{3+}$ ion's spin state and can be considered spin-related bands in the SERS spectrum. Figure 5(d) shows a comparison of the scaled simulated SERS spectrum and the experimental spectrum under 532 nm laser excitation of Gd-citrate. Although there were some differences between the simulated and experimental spectra, the intensity of the $\nu$(C-O-Gd) was lower than that of $\nu_{sym}(COO^-)$, while the intensity of the $\nu_{sym}(COO^-)$ was stronger than that of the $\nu_{asym}(COO^-)$ both in simulated and experimental spectra. Thus, this relative intensity feature, which may relate to the spin state, was similar to the experimental SERS spectrum excited at 532 nm and was also exhibited by the DFT calculation results. It is possibly caused the difference in the electron density in the same area of the RE-citrate complex. In Figure 5(a) and (c), we compared the electronic structures of citrate regions around different $RE^{3+}$ ions by spin population. The symmetry relationship of the electronic structure was roughly shown in the spin density. The Gd-citrate having 7 more single electrons (alpha electrons) than La-citrate resulting in spin density of it was not 0. Especially at the hydroxyl group where citrate coordinates with $Gd^{3+}$ ions, indicating that the electron density in this part was higher. Hence, the polarization of C-O-Gd is comparatively more arduous than that of C-O-La when subjected to laser light of identical wavelengths, thereby leading to a reduced Raman intensity. In addition, we also considered the mass effect of ions. For the RE-citrate complexes, the frequencies of $COO^-$ coordinated to $RE^{3+}$ ions, e.g. near 940, 1315 and 1495 $cm^{-1}$, are significantly lower than the frequencies of $COO^-$ for citrate in Table 2, e.g. at 956, 1390 and 1580 $cm^{-1}$. This is due to the fact that La and Gd atoms are much heavier than H atoms. In contrast, comparing the SERS spectra of the experimental RE-citrate complexes, the mass effect is not significant. It may be due to the vibrational contribution of $RE^{3+}$ were too weak, which studied by DFT simulations.

Such experimental and computational results indicate that a large spin difference between the $RE^{3+}$ ions, $La^{3+}$ and $Gd^{3+}$, can be observed in the relative intensities of the characteristic peaks of their SERS spectra. When $RE^{3+}$ ions coordinate with the hydroxyl and carboxyl groups of citrate molecules, the corresponding SERS characteristic peaks are around 1065 cm$^{-1}$ and 1315 cm$^{-1}$, and can be considered as spin-related bands. The presence of these spin-related bands in the SERS spectrum could be used for marking the spin state of the complex based on the relative intensity relationship of spin-related peaks. This method is similar to that reported by Kitahama et al[12]. The relative intensity ratio of the two spin-related bands is not affected by the laser wavelength, thus it can be used to classify La$^{3+}$ and Gd$^{3+}$ ions based on their SERS spectra.

**Conclusion**

In this study, we found that the SERS spectra of RE-citrate complexes (RE: La, Gd) could be used for the qualitative classification of La$^{3+}$ and Gd$^{3+}$ ions, possibly because of the spin-related bands in the SERS spectrum. By evaluating the spectral relationship of sample solutions with and without citrate@AgNPs, we confirmed that the SERS spectrum of RE-citrate complexes could be obtained by adding a small amount of RE$^{3+}$ solution to citrate@AgNPs. In the SERS spectra, the positions of the SERS characteristic peaks of La-citrate and Gd-citrate were similar, and these peak positions did not change significantly with the change in excitation wavelength. However, the intensities of the SERS characteristic peaks were different, especially in the interval from 1000 to 1800 cm$^{-1}$. Three characteristic peaks around 1065, 1315, and 1495 cm$^{-1}$ were used to explore the differences in the SERS spectra of the different RE-citrate complexes. Combined with DFT calculations, the characteristic SERS peaks near 1065 cm$^{-1}$ and 1315 cm$^{-1}$ were assigned to spin-related bands (C-O-RE) and (COO$^-$), respectively. Their relative intensity changes were related to the electron density in the region affected by RE$^{3+}$ ions. Moreover, the relative intensity relationship of the SERS peaks near 1065 cm$^{-1}$ and 1315 cm$^{-1}$ can be used to classify La$^{3+}$ and Gd$^{3+}$ ions. The ratios of $I_{1065}/I_{1315}$ for La-citrate and Gd-citrate were in the range of approximately 1 and 0.5, respectively. The results of the present study show that using the physical properties of RE elements, such as the spin state, RE-citrate complexes can be constructed as target molecules to study the differences in their SERS spectra, thus enabling the classification of different RE$^{3+}$ ions via SERS, as well as examining the effects of RE$^{3+}$ ions on SERS. In future studies, we will investigate the effects of the spin state of other RE$^{3+}$ ions, such as Pr$^{3+}$ and Nd$^{3+}$ ions, on the SERS spectra, and DFT SERS simulations of RE-molecule complexes.

**Acknowledgments**

The authors acknowledge Dr. Lu Tian (Beijing Kein Research Center for Natural Sciences, China) for his fruitful suggestions in DFT calculation of rare earth elements. The authors acknowledge funding from JSPS KAKENHI Grant-in-Aid for Scientific Research (C), number 21K04935.

**References**

(1) Rainer Pöttgen; Jüstel, T.; Strassert, C. A. *Rare Earth Chemistry*; Walter de Gruyter GmbH & Co KG, 2020.

(2) Eliseeva, S. V.; Bünzli, J.-C. G. Lanthanide Luminescence for Functional Materials and Bio-Sciences. *Chem. Soc. Rev.* **2010**, 39 (1), 189-227.

(3) Wang, G.; Peng, Q.; Li, Y. Lanthanide-Doped Nanocrystals: Synthesis, Optical-Magnetic Properties, and Applications. *Accounts of Chemical Research* **2011**, 44 (5), 322–332.

(4) Huang, C.-H. *Rare Earth Coordination Chemistry*; John Wiley & Sons, 2011.

(5) Kotyk, C. M.; Weber, J. E.; Hyre, A. S.; McNeely, J.; Monteiro, J. H. S. K.; Domin, M.; Balaich, G. J.; Rheingold, A. L.; de Bettencourt-Dias, A.; Doerrer, L. H. Luminescence of Lanthanide Complexes with Perfluorinated Alkoxide Ligands. *Inorganic Chemistry* **2020**, 59 (14), 9807–9823.

(6) Lee, H. Y.; Jee, H. W.; Seo, S. M.; Kwak, B. K.; Khang, G.; Cho, S. H. Diethylenetriaminepentaacetic Acid−Gadolinium (DTPA-Gd)-Conjugated Polysuccinimide Derivatives as Magnetic Resonance Imaging Contrast Agents. *Bioconjugate Chemistry* **2006**, *17* (3), 700–706.

(7) Martínez-Pérez, M. J.; Cardona-Serra, S.; Schlegel, C.; Moro, F.; Alonso, P. J.; Prima-

García, H.; Clemente-Juan, J. M.; Evangelisti, M.; Gaita-Ariño, A.; Sesé, J.; van Slageren, J.; Coronado, E.; Luis, F. Gd-Based Single-Ion Magnets with Tunable Magnetic Anisotropy: Molecular Design of Spin Qubits. *Physical Review Letters* **2012**, 108 (24).

(8) Skripka, A.; Karabanovas, V.; Jarockyte, G.; Marin, R.; Tam, V.; Cerruti, M.; Rotomskis, R.; Vetrone, F. Theranostics: Decoupling Theranostics with Rare Earth Doped Nanoparticles (Adv. Funct. Mater. 12/2019). *Advanced Functional Materials* **2019**, 29 (12), 1970073.

(9) Serrano, D.; Kuppusamy, S. K.; Heinrich, B.; Fuhr, O.; Hunger, D.; Ruben, M.; Goldner, P. Ultra-Narrow Optical Linewidths in Rare-Earth Molecular Crystals. *Nature* **2022**, 603 (7900), 241–246.

(10) Liu, Y.; Chang, Z.; Yuan, H.; Fales, A. M.; Vo-Dinh, T. Quintuple-Modality (SERS-MRI-CT-TPL-PTT) Plasmonic Nanoprobe for Theranostics. *Nanoscale* **2013**, 5 (24), 12126.

(11) Colthup, N. B.; Daly, L. H.; Wiberley, S. E. *Introduction to Infrared and Raman Spectroscopy*; Academic Press: San Diego, Calif., 1998.

(12) Kitahama, Y.; Egashira, M.; Suzuki, T.; Tanabe, I.; Ozaki, Y. Sensitive Marker Bands for the Detection of Spin States of Heme in Surface-Enhanced Resonance Raman Scattering Spectra of Metmyoglobin. *The Analyst* **2014**, 139 (24), 6421–6425.

(13) Ozaki, Y.; Kitagawa, T.; Kyogoku, Y. Raman Study of the Acid-Base Transition of Ferric Myoglobin; Direct Evidence for the Existence of Two Molecular Species at Alkaline PH. *FEBS Letters* **1976**, 62 (3), 369–372.

(14) Nie, S. Probing Single Molecules and Single Nanoparticles by Surface-Enhanced Raman Scattering. *Science* **1997**, 275 (5303), 1102–1106.

(15) Fleischmann, M.; Hendra, P. J.; McQuillan, A. J. Raman Spectra of Pyridine Adsorbed at a Silver Electrode. *Chemical Physics Letters* **1974**, 26 (2), 163–166.

(16) Albrecht, M. G.; Creighton, J. A. Anomalously Intense Raman Spectra of Pyridine at a Silver Electrode. *J. Am. Chem. Soc*. **1977**, 99, 5215−5217.

(17) Jeanmaire, D. L.; Van Duyne, R. P. Surface Raman Spectroelectrochemistry: Part I. Heterocyclic, Aromatic, and Aliphatic Amines Adsorbed on the Anodized Silver Electrode. *J. Electroanal.Chem*. **1977**, 84, 1−20.

(18) Link, S.; El-Sayed, M. A. Spectral Properties and Relaxation Dynamics of Surface Plasmon Electronic Oscillations in Gold and Silver Nanodots and Nanorods. *The Journal of Physical Chemistry B* **1999**, 103 (40), 8410–8426.

(19) Yamamoto, Y. S.; Itoh, T. Why and How Do the Shapes of Surface-Enhanced Raman Scattering Spectra Change? Recent Progress from Mechanistic Studies. *Journal of Raman Spectroscopy* **2016**, *47* (1), 78–88.

(20) Itoh, T.; Procházka, M.; Dong, Z.-C.; Ji, W.; Yamamoto, Y. S.; Zhang, Y.; Ozaki, Y. Toward a New Era of SERS and TERS at the Nanometer Scale: From Fundamentals to Innovative Applications. *Chemical Reviews* **2023**, *123* (4), 1552–1634.

(21) Yang, S.; Yao, J.; Quan, Y.; Hu, M.; Su, R.; Gao, M.; Han, D.; Yang, J. Monitoring the Charge-Transfer Process in a Nd-Doped Semiconductor Based on Photoluminescence and SERS Technology. *Light: Science & Applications* **2020**, 9 (1).

(22) Runowski, M.; Goderski, S.; Paczesny, J.; Księżopolska-Gocalska, M.; Ekner-Grzyb, A.; Grzyb, T.; Rybka, J. D.; Giersig, M.; Lis, S. Preparation of Biocompatible, Luminescent-Plasmonic Core/Shell Nanomaterials Based on Lanthanide and Gold Nanoparticles Exhibiting SERS Effects. *The Journal of Physical Chemistry C* **2016**, 120 (41), 23788–23798.

(23) Derom, S.; Berthelot, A.; Pillonnet, A.; Benamara, O.; Jurdyc, A. M.; Girard, C.; Colas des Francs, G. Metal Enhanced Fluorescence in Rare Earth Doped Plasmonic Core–Shell Nanoparticles. *Nanotechnology* **2013**, 24 (49), 495704.

(24) Wang, J.; Huang, H.; Zhang, D.; Chen, M.; Zhang, Y.; Yu, X.; Zhou, L.; Wang, Q. Synthesis of Gold/Rare-Earth-Vanadate Core/Shell Nanorods for Integrating Plasmon Resonance and Fluorescence. *Nano Research* **2015**, 8 (8), 2548–2561.

(25) López-Neira, J. P.; Galicia-Hernández, J. M.; Reyes-Coronado, A.; Pérez, E.;

Castillo-Rivera, F. Correction to "Surface Enhanced Raman Scattering of Amino Acids Assisted by Gold Nanoparticles and $Gd^{3+}$ Ions." *The Journal of Physical Chemistry A* **2015**, *119* (22), 5900–5900.

(26) Carnall, W. T. The Absorption and Fluorescence Spectra of Rare Earth Ions in Solution. Handbook on the Physics and Chemistry of Rare Earths, Vol. 3, 171-208, 1979.

(27) Munro, C. H.; Smith, W. E.; Garner, M.; Clarkson, J.; White, P. C. Characterization of the Surface of a Citrate-Reduced Colloid Optimized for Use as a Substrate for Surface-Enhanced Resonance Raman Scattering. *Langmuir* **1995**, *11* (10), 3712–3720.

(28) Lee, P. C.; Meisel, D. Adsorption and Surface-Enhanced Raman of Dyes on Silver and Gold Sols. *The Journal of Physical Chemistry* **1982**, *86* (17), 3391–3395.

(29) Ivanova, V. Yu.; Shurygin, I. D.; Chevela, V. V.; Ajsuvakova, O. P.; Semenov, V. E.; Bezryadin, S. G. New Aspects of Complex Formation in the Gadolinium(III)–Citric Acid System in Aqueous Solution. *Comments on Inorganic Chemistry* **2021**, *42* (2), 109–144.

(30) Grimme, S.; Antony, J.; Ehrlich, S.; Krieg, H. A Consistent and Accurate Ab Initio Parametrization of Density Functional Dispersion Correction (DFT-D) for the 94 Elements H-Pu. *The Journal of Chemical Physics* **2010**, *132* (15), 154104.

(31) Paier, J.; Marsman, M.; Kresse, G. Why Does the B3LYP Hybrid Functional Fail for Metals? *The Journal of Chemical Physics* **2007**, *127* (2), 024103.

(32) Adamo, C.; Barone, V. Toward Reliable Density Functional Methods without Adjustable Parameters: The PBE0 Model. *The Journal of Chemical Physics* **1999**, *110* (13), 6158–6170.

(33) Grimme, S.; Antony, J.; Ehrlich, S.; Krieg, H. A Consistent and Accurate Ab Initio Parametrization of Density Functional Dispersion Correction (DFT-D) for the 94 Elements H-Pu. *The Journal of Chemical Physics* **2010**, *132* (15), 154104.

(34) Chen, X.; Chen, T.-T.; Li, W.-L.; Lu, J.-B.; Zhao, L.-J.; Jian, T.; Hu, H.-S.; Wang, L.-S.; Li, J. Lanthanides with Unusually Low Oxidation States in the PrB3- and PrB4-Boride Clusters. *Inorganic Chemistry* **2019**, *58* (1), 411–418.

(35) Brémond, É.; Savarese, M.; Su, N. Q.; Pérez-Jiménez, Á. J.; Xu, X.; Sancho-García, J. C.; Adamo, C. Benchmarking Density Functionals on Structural Parameters of Small-/Medium-Sized Organic Molecules. *Journal of Chemical Theory and Computation* **2016**, *12* (2), 459–465.

(36) Dunning, T. H. Gaussian Basis Sets for Use in Correlated Molecular Calculations. I. The Atoms Boron through Neon and Hydrogen. *The Journal of Chemical Physics* **1989**, *90* (2), 1007–1023.

(37) Andrae, D.; Häußermann U.; Dolg, M.; Stoll, H.; Preuß, H. Energy-Adjustedab Initio Pseudopotentials for the Second and Third Row Transition Elements. *Theoretica Chimica*

*Acta* **1990**, *77* (2), 123–141.

(38) Dolg, M.; Stoll, H.; Savin, A.; Preuss, H. Energy-Adjusted Pseudopotentials for the Rare Earth Elements. *Theoretica Chimica Acta* **1989**, *75* (3), 173–194.

(39) Dolg, M.; Stoll, H.; Preuss, H. A Combination of Quasirelativistic Pseudopotential and Ligand Field Calculations for Lanthanoid Compounds. *Theoretica Chimica Acta* **1993**, *85* (6), 441–450.

(40) Rappoport, D. Property-Optimized Gaussian Basis Sets for Lanthanides. *The Journal of Chemical Physics* **2021**, *155* (12), 124102.

(41) Kashinski, D. O.; Chase, G. M.; Nelson, R. G.; Di Nallo, O. E.; Scales, A. N.; VanderLey, D. L.; Byrd, E. F. C. Harmonic Vibrational Frequencies: Approximate Global Scaling Factors for TPSS, M06, and M11 Functional Families Using Several Common Basis Sets. *The Journal of Physical Chemistry. A* **2017**, *121* (11), 2265–2273.

(42) Kesharwani, M. K.; Brauer, B.; Martin, J. M. L. Frequency and Zero-Point Vibrational Energy Scale Factors for Double-Hybrid Density Functionals (and Other Selected Methods): Can Anharmonic Force Fields Be Avoided? *The Journal of Physical Chemistry A* **2014**, *119* (9), 1701–1714.

(43) Weigend, F.; Ahlrichs, R. Balanced Basis Sets of Split Valence, Triple Zeta Valence and Quadruple Zeta Valence Quality for H to Rn: Design and Assessment of Accuracy. *Physical Chemistry Chemical Physics* **2005**, *7* (18), 3297.

(44) Lu, T.; Chen, F. Multiwfn: A Multifunctional Wavefunction Analyzer. *Journal of Computational Chemistry* **2011**, *33* (5), 580–592.

(45) Humphrey, W.; Dalke, A.; Schulten, K. VMD: Visual Molecular Dynamics. *Journal of Molecular Graphics* **1996**, *14* (1), 33–38.

(46) Park, J.-W.; Shumaker-Parry, J. S. Structural Study of Citrate Layers on Gold Nanoparticles: Role of Intermolecular Interactions in Stabilizing Nanoparticles. *Journal of the American Chemical Society* **2014**, *136* (5), 1907–1921.

(47) Ahuja, T.; Chaudhari, K.; Paramasivam, G.; Ragupathy, G.; Mohanty, J. S.; Pradeep, T. Toward Vibrational Tomography of Citrate on Dynamically Changing Individual Silver Nanoparticles. *The Journal of Physical Chemistry C* **2021**, *125* (6), 3553–3566.

(48) Vanhoyland, G.; Pagnaer, J.; D'Haen, J.; Mullens, S.; Mullens, J. Characterization and Structural Study of Lanthanum Citrate Trihydrate [La(C6H5O7)(H2O)2]·H2O. *Journal of Solid State Chemistry* **2005**, *178* (1), 166–171.

(49) Yon, M.; Pibourret, C.; Marty, J.-D.; Ciuculescu-Pradines, D. Easy Colorimetric Detection of Gadolinium Ions Based on Gold Nanoparticles: Key Role of Phosphine-Sulfonate Ligands. *Nanoscale Advances* **2020**, *2* (10), 4671–4681.

(50) Yamamoto, Y. S.; Ozaki, Y.; Itoh, T. Recent Progress and Frontiers in the

Electromagnetic Mechanism of Surface-Enhanced Raman Scattering. *Journal of Photochemistry and Photobiology C: Photochemistry Reviews* **2014**, *21*, 81–104.

(51) Xu, H.; Wang, X.-H.; Persson, M. P.; Xu, H. Q.; Käll, M.; Johansson, P. Unified Treatment of Fluorescence and Raman Scattering Processes near Metal Surfaces. *Physical Review Letters* **2004**, *93* (24).

(52) Kerker, M.; Siiman, O.; Bumm, L. A.; Wang, D.-S. . Surface Enhanced Raman Scattering (SERS) of Citrate Ion Adsorbed on Colloidal Silver. *Applied Optics* **1980**, *19* (19), 3253.

(53) Nakabayashi, T.; Kosugi, K.; Nishi, N. Liquid Structure of Acetic Acid Studied by Raman Spectroscopy and Ab Initio Molecular Orbital Calculations. *The Journal of Physical Chemistry A* **1999**, *103* (43), 8595–8603.

(54) Liu, Z.; Lu, T.; Chen, Q. Vibrational Spectra and Molecular Vibrational Behaviors of All-Carboatomic Rings, Cyclo[18]Carbon and Its Analogues. *Chemistry – An Asian Journal* **2020**, *16* (1), 56–63.

(55) Parr, R. G.; Yang, W. *Density-Functional Theory of Atoms and Molecules*; New York, Ny Oxford Univ. Press, 1994.

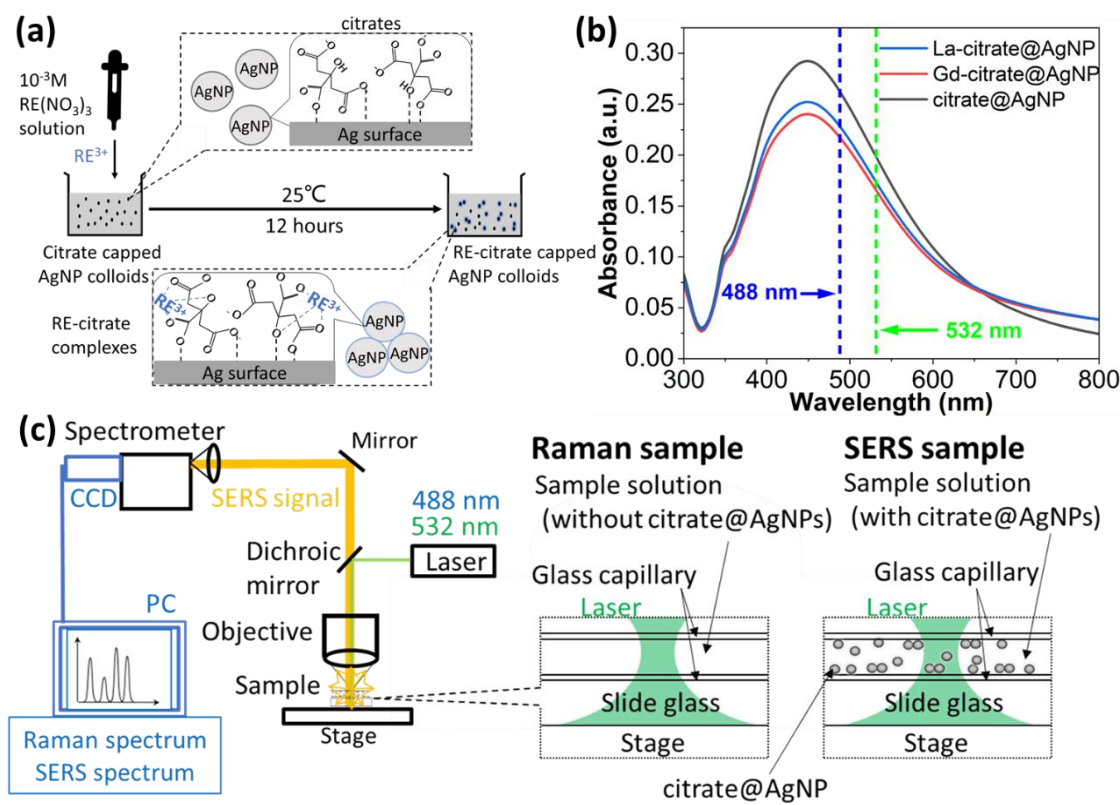

Figure 1. (a) A schematic representation of the interaction of $RE^{3+}$ (RE: La, Gd) ions with the surfaces of citrate-capped sliver nanoparticles (citrate@AgNPs). (b) UV-Visible absorption spectra of La-citrate@AgNPs, Gd-citrate@AgNPs and citrate@AgNPs. (c) A schematic of measurement system for Raman and SERS spectroscopy. Raman and SERS samples were enclosed in a glass capillary, and then their Raman and SERS spectra were measured using 488 nm and 532 nm laser lights, respectively. Here, we defined that Raman samples are solutions without citrate@AgNPs, and SERS samples are those with citrate@AgNP, respectively.

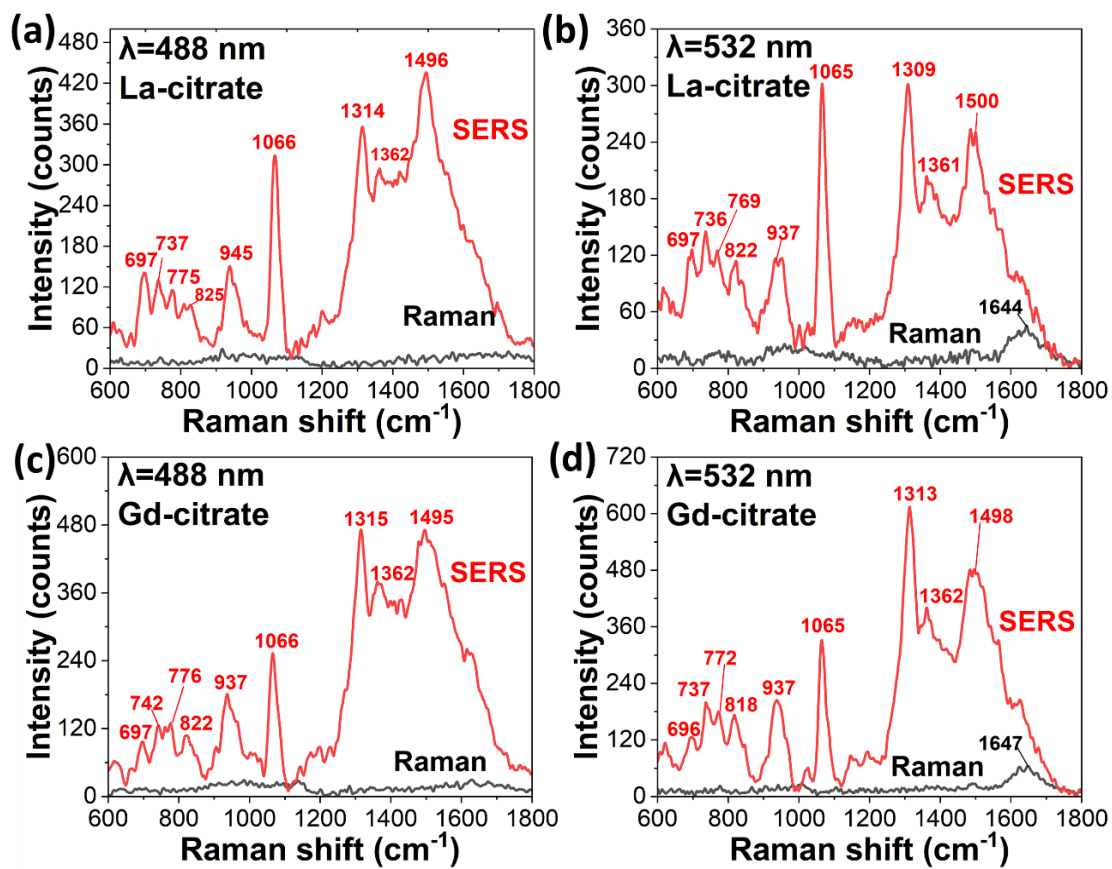

Figure 2. Raman (black line) and SERS (red line) spectra of (a) La-citrate excited at 488 nm, (b) La-citrate excited at 532 nm, (c) Gd-citrate excited at 488 nm and (d) Gd-citrate excited at 532 nm.

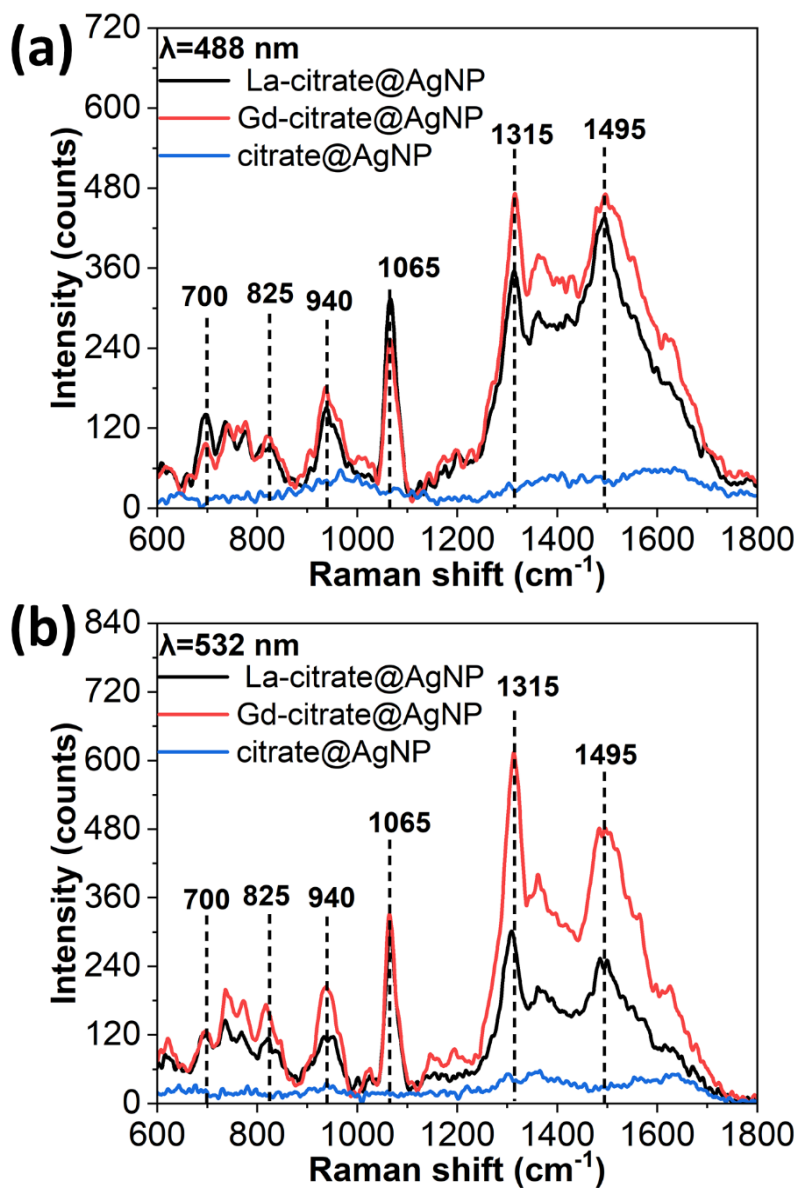

Figure 3. SERS spectra of La-citrate (black line), Gd-citrate (red line) and citrate (blue line) excited at (a) 488 nm and (b) 532 nm. Peak assignments are shown in Table 1.

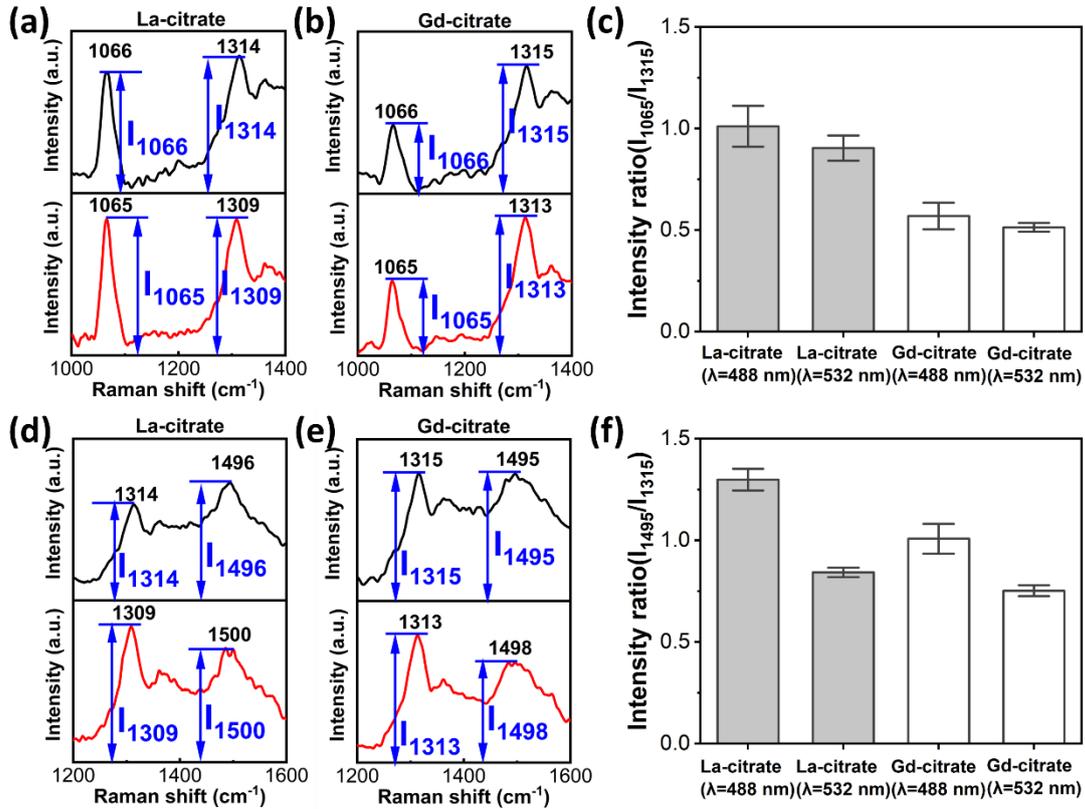

Figure 4. (a) and (b): SERS spectra in the region of 1000 − 1400 cm$^{-1}$ excited at 488 nm (black line) and 532 nm (red line) of (a) La-citrate and (b) Gd-citrate. The blue line shows the method to obtain the intensities of characteristic peaks. (c): Ratios of intensities of characteristic peaks around 1065 cm$^{-1}$ and 1315 cm$^{-1}$ under different wavelength lasers' excitations for ion classification. (d) and (e): SERS spectra in the region of 1200 − 1600 cm$^{-1}$ excited at 488 nm (black line) and 532 nm (red line) of La-citrate (d) and Gd-citrate (e). The blue line shows the method to obtain the intensities of characteristic peaks. (f): Ratios of intensities of characteristic peaks around 1315 cm$^{-1}$ and 1495 cm$^{-1}$ under different wavelength lasers' excitations for ion classification. For (c) and (f), the numbers

of data analyzed were 9, 8, 9 and 8 for La-citrate (488 nm), La-citrate (532 nm), Gd-citrate (488 nm) and Gd-citrate (532 nm), respectively. Error bars are ± SD.

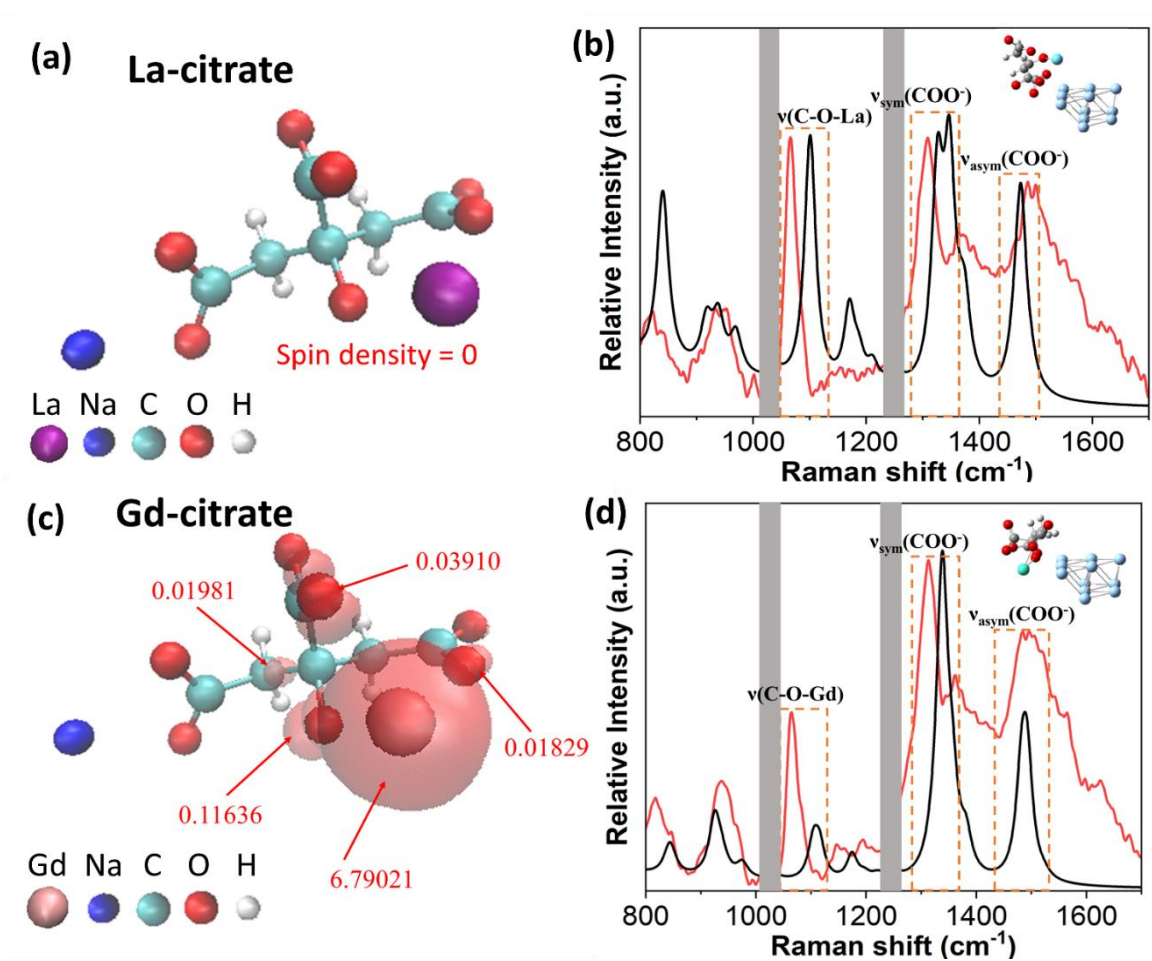

Figure 5. (a) Structure and spin population of La-citrate. (b) Experimental SERS spectrum excited at 532 nm (red line), simulated SERS spectrum (black line) and an estimated structure of La-citrate adsorbed on silver surface. (c) Structure and spin population of Gd-citrate. (d) Experimental SERS spectrum excited at 532 nm (red line), simulated SERS spectrum (black line) and an estimated structure of Gd-citrate adsorbed on silver surface. All simulated SERS spectra were scaled using scaled factors.

Table 1. Peak assignment for SERS spectrum of La-citrate and Gd-citrate

| SERS of La-citrate | | | SERS of Gd-citrate | | |
|---|---|---|---|---|---|
| Raman shift (cm$^{-1}$) | | Assignment | Raman shift (cm$^{-1}$) | | Assignment |
| Experimental | Simulated | | Experimental | Simulated | |
| 617, 697, | 629, 681, | $\delta$(COO$^-$) | 622, 696, | 635, 684, | $\delta$(COO$^-$) |
| 736, 737, 769, 775 | 712, 745 | $\delta$(COO$^-$), $\gamma$(COO$^-$) | 737, 742, 776, 772 | 716, 747 | $\delta$(COO$^-$), $\gamma$(COO$^-$) |
| 822, 825 | 840 | $\nu$(CCCC-O) | 818, 822 | 844 | $\nu$(CCCC-O) |
| 937，945 | 920, 938, 973 | $\nu$(C-COO$^-$), $\delta$(CH$_2$) | 937 | 928, 975 | $\nu$(C-COO$^-$), $\delta$(CH$_2$) |
| 1065，1066 | 1100 | $\nu$(C-O-La), $\gamma$(CH$_2$) | 1064，1067 | 1109 | $\nu$(C-O-Gd), $\gamma$(CH$_2$) |
| 1309，1314 | 1327, 1344 | $\nu_{sym}$(COO$^-$), $\delta$(CH$_2$) | 1313，1315 | 1339 | $\nu_{sym}$(COO$^-$), $\delta$(CH$_2$) |
| 1496, 1500 | 1473 | $\nu_{asym}$(COO$^-$), $\gamma$(CH$_2$) | 1495, 1498 | 1488 | $\nu_{asym}$(COO$^-$), $\gamma$(CH$_2$) |

Notes: $\nu$ indicates stretching, $\nu_{sym}$ is symmetric stretching, $\nu_{asym}$ is asymmetric stretching, $\delta$ is in-plane bending and rocking, and $\gamma$ is out-of-plane wagging and twisting. The calculated SERS frequencies were scaled by using scaled factors.

Table 2. Peak assignment for SERS spectrum of citrate[27] [52]

| Raman Shift (cm$^{-1}$) | Assignment |
| --- | --- |
| 670[27] | $\delta(COO^-)$ |
| 845[52] | $\nu(CCCC-O)$, |
| 933[27], 956[52] | $\nu(C-COO^-)$ |
| 1025[27] | $\nu(C-OH)$ |
| 1390[27], 1400 [52], 1415[52] | $\nu_{sym}(COO^-)$ |
| 1580[52] | $\nu_{asym}(COO^-)$ |

Notes: $\nu$ indicates stretching, $\nu_{sym}$ is symmetric stretching, $\nu_{asym}$ is asymmetric stretching, $\delta$ is in-plane bending and rocking.